\documentclass[runningheads]{llncs}
\usepackage[T1]{fontenc}
\usepackage{graphicx}
\usepackage{url,color}

\usepackage{multirow}
\usepackage{wrapfig}
\usepackage{booktabs}
\usepackage{array}
\usepackage{tikz}
\usepackage{adjustbox}
\usepackage{amsmath}
\usepackage{subcaption}
\usepackage{floatrow}
\usepackage{longtable}
\usepackage{amssymb} 
\usepackage{pgfplots}
\usepackage{pgfplotstable}
\pgfplotsset{compat=1.18}
\definecolor{seccol}{rgb}{0.05, 0.15, 0.7} 

\usepackage{tabularx}

\begin{document}

\title{Evaluating ML-based Intrusion Detection Systems: The Illusion of Model Efficacy}
%
%
\author{Achilleas Spanos\inst{1} \and
Ioanna Kantzavelou\inst{1}\orcidID{0000-0001-5976-7837}}
%
\titlerunning{The Illusion of Model Efficacy}
\authorrunning{A. Spanos et al.}
%
\institute{University of West Attica, Egaleo, Greece}

%
\maketitle              
\begin{abstract}

Intrusion Detection has been revolutionized due to the integration of Machine Learning (ML). Improved detection rate, reduced false alarms, and optimized algorithms contribute to the perception of improved systems with optimal accuracy and near-perfect performance, \textit{the illusion of model efficacy}. However, the value of this effectiveness diminishes when confronted with unseen attacks. In this paper, we go beyond solely algorithmic enhancements and metric adjustments in ML-based Network Intrusion Detection Systems. We design an experiment to test the generalization capabilities of certain classifiers on unseen attacks. Our approach examines the dimensionality parameter’s impact through two experimental methodologies, which are applied in two distinct settings. The experimental findings reveal how effectively the models could identify even a fraction of unseen attacks and underscore structural weaknesses in ML-based IDS research and evaluation techniques. Finally, seven evaluation criteria are outlined to address these challenges.

\end{abstract}
\section{Introduction}

In the absence of effective security mechanisms that promptly restrict synchronous attacks, the area of Intrusion Detection has eventually stabilized a crucial role within Cybersecurity~\cite{hozouri2025comprehensive}. In recent years, it is Machine Learning that has revolutionized Intrusion Detection even more. In such ML-based studies,  most researchers aim to enhance the accuracy of a detection engine \cite{talukder2023dependable}, improve scalability \cite{talukder2024machine} and interpretability \cite{wei2023xnids},  optimize rule generalization \cite{coscia2024automatic}, increase zero-day attack detection \cite{kumar2021robust}, and strengthen its resilience to adversary techniques \cite{jmila2022adversarial}. Only a small fraction of these research works attempt to address challenges that bridge research with real-world deployments.

Many distinct frameworks and methodologies that have been proposed prioritize different models, preprocessing techniques \cite{umar2025effects}, dimensionality reduction techniques \cite{abdulhammed2019features}, and data types \cite{liu2019machine}. However, only a handful of research studies focus on the real challenges in Network Intrusion Detection Systems (NIDS) that expand beyond slightly algorithmic improvements and evaluation metrics. In particular, high accuracy against attacks within the training distribution is not an indication of strong detection performance, and although it can be improved, it is already well developed.

Current literature is shifting beyond the prominent problems in network intrusion detection. Instead of focusing on zero-day robustness, interpretability, attack patterns, and relations, adversarial robustness and multimodal techniques, more and more research works propose frameworks that are evaluated on in-training-distribution data and known attacks, aiming at slight algorithmic and metric improvements.

The proposed research work attempts to address real-world problems of ML-based Network Intrusion Detection Systems. We argue that an explainable and generalized IDS would outperform any \textit{"optimal"} presented IDS despite the potential tradeoffs in overall accuracy. To address this, and considering the extensive amount of research work as a motive, we design a simple yet meaningful experiment to test the generalization capabilities of classifiers, while assessing all the underlying and hidden risks it may exhibit. Additionally, we consolidate a set of evaluation criteria for ML-based NIDS, reframing the research gap around security necessities rather than benchmark metrics.


Contributions: The contributions of this paper are pointed out in the sequel:
\begin{itemize}
    \item We delve into zero-day robustness, instead of evaluating on in-training-distribution data and known attacks.
    \item We design a simple and meaningful experiment to test the generalization capabilities of certain classifiers.   
    \item We assess what the experiment reveals as underlying and hidden risks that might exist in an \textit{”ideal”} ML-based network IDS.
    \item  We explore a proposed attack-based intra-dataset generalization evaluation, providing a high-level model’s generalization evaluation without introducing domain shifts.
    \item We employ three feature selection approaches to examine the dimensionality parameter in the generalization behavior of classifiers, design two experimental methodologies, and apply them in two experimental settings.
    \item We consolidate a set of evaluation criteria for ML-based NIDS, reframing the research gap around security necessities rather than benchmark metrics.
\end{itemize}

The remainder of this paper is organized as follows: Section~\ref{sec:relatedwork} presents the existing work in the literature with respect to both the experimental study and the evaluation criteria for a NIDS. Section~\ref{sec:methodology} demonstrates the two experimental setups, methodologies, configurations and scopes. Section~\ref{sec:results} concerns a metric-score comparison of the selected components and parameters. In Section~\ref{sec:findings} the experimental findings are translated into actionable concerns and blind spots that exist in the literature. Systematic evaluation criteria are presented in Section~\ref{sec:evaluation}, extending towards the enhancement of generalized intrusion detection and continual research.

\section{Background and Related Work}
\label{sec:relatedwork}

ML-based NIDS research has consistently demonstrated optimal detection on public benchmarks.  However, only a limited number of studies have raised concerns about whether these scores accurately reflect the operational capabilities of such systems. This section presents two distinct strands of literature. The first, \textit{the Generalization Gap} empirically examines the behavior of ML models on Out-Of-Distribution samples, consistently observing non-transferable performance. The second, \textit{the Evaluation Practices} critiques the evaluation practices employed in ML-security research, identifying methodological pitfalls and proposing standards for the development and evaluation of ML-based NIDS.

\subsection{Generalization Gap}\label{sec:first}

The operational performance of ML-based NIDS has been frequently defined as poor generalization beyond the training distribution. In 2010, it was first argued that the inherent characteristics of intrusion detection conflict with the closed-world ML assumptions~\cite{sommer2010outside}. This is because deploying an ML-based NIDS exposes the target system to an adversarial setting. Subsequent empirical work has quantified this fragility through cross-dataset experimental settings.

Considering the structural similarity between CICIDS2017 and CSE-CIC-IDS2018, twelve supervised classifiers were trained on one dataset and evaluated on the other. The results indicated that classifiers failed to detect unseen network traffic, even when the evaluation was restricted to the same attack classes~\cite{d2020inter}. A similar generalization collapse for unsupervised models has been reported~\cite{verkerken2022towards}, which justifies the lack of real-world adoption of anomaly-based NIDS. Apruzzese et al.~\cite{apruzzese2022cross} further formalize cross-dataset evaluation over six datasets to address the specificity issue and reveal the true potential of generalized NIDS. 

In contrast, Cantone et al.~\cite{cantone2024machine} demonstrated on four datasets that near-perfect within-dataset scores does not generalize across datasets, as there are no isolated attack–dataset pairs. In an effort to enhance rather than diagnose the cross-dataset transferability gap, Sudyana et al. introduce a lifecycle-based dataset and auto-learning features, thereby demonstrating the generalization impact.

All of these studies employ similar methodologies, with varying network environments between training and testing. This indicates that the reported drops in performance are attributable to domain shift and the novelty of attacks. The few studies that maintain the network environments identical employ Leave-One-Attack-Out protocols, where a single attack is excluded at a time~(\cite{chettri2026analyzing},~\cite{uddin2024usfad},~\cite{zoppi2023algorithm}). Similarly, other studies train anomaly detectors on benign samples and evaluate per-attack holdout~\cite{hindy2020utilising}. However, such approaches are narrow and do not accurately reflect the simultaneous and multi-attack shift a detector encounters.

In our research work, we keep the network environment intact and examine the generalization at an advanced level. The selected approach enables us to uncover the potential for exploiting common attack patterns and conducting comprehensive generalization testing.

\subsection{Evaluation Practices}

Beyond the empirical generalization assessment, research studies transformed critical gaps into evaluation practices, shortcomings, dataset auditing, and ML security research guidelines. At the broadest level, Arp et al.~\cite{arp2022and} identify ten recurring pitfalls that distort reported performance across security-ML studies and propose actionable recommendations beyond generic benchmarks. In the network intrusion detection research domain, the gap between research results and practitioner trust has been systematically addressed~\cite{apruzzese2023sok}, introducing the concept of pragmatic assessment and re-assessing the evaluation methodology to estimate the real quality of ML-based NIDS.

Numerous research studies have consolidated on the influence of datasets on the reported results and findings of NIDS~\cite{thakkar2020review}. Labeling errors, flow-construction and feature-extraction flaws, as well as artificial temporal separation have been documented as severe defects in CIC-IDS2017 and CSE-CIC-IDS2018 datasets. These defects could substantially alter the reported detection output (\cite{engelen2021troubleshooting}, \cite{lanvin2022errors}, \cite{liu2022error}). The suitability of datasets has been repeatedly criticized in the literature. Gharib et al.~\cite{gharib2016evaluation} assess criteria for judging the quality of IDS benchmark datasets, while some other studies question whether headline gains on these public datasets amount to real advances at all~\cite{catillo2023machine}.

None of the research studies derive a set of evaluation criteria from an empirical generalization failure, nor pair each criterion with a concrete reporting artifact. Building upon the literature and motivated by our empirical findings presented in Section~\ref{sec:findings}, we consolidate a set of evaluation criteria for ML-based NIDS. By adopting these evaluation criteria the model remains intact, operating solely as an additional layer of reporting. The selected approach reframes the research gap around security necessities rather than benchmark metrics.

\section{Experimental Methodology}
\label{sec:methodology}

The experimental methodology presented in this section was designed to assess, at a high level, the ability of ML classifiers to generalize to unseen attacks. The core methodology is established on cross-attack intra-dataset evaluation to isolate the effect of unseen attacks in a controlled, reproducible, and realistic generalization test. Subsection~\ref{subsec:dataset} will provide justification for the dataset selection, which serves as the core experimental setup. Subsection~\ref{subsec:design} outlines the core experimental methodology, encompassing three feature-selection strategies: \textit{Full-feature-set}, \textit{Local}, and \textit{Greedy-global}. These strategies have been evaluated through two distinct experimental testbeds.

\subsection{Dataset Selection}
\label{subsec:dataset}

The process of dataset selection determines the attack coverage, enables results comparison, but also operates as a methodological motive for the experimental study. A plethora of intrusion detection datasets have been studied, considering factors such as collection methodology,  attack diversity, traffic characteristics, and adoption frequency within the ML-based IDS literature. An extensive review of the NIDS datasets, which is beyond the scope of this work, can be found in Goldschmidt et al. research study \cite{goldschmidt2025network}. Due to the balance between recency and broad adoption within the research community\cite{goldschmidt2025network}, the CICIDS2017~\cite{sharafaldin2018toward} dataset has been selected. The dataset's dual-level structure includes both raw packet captures and flow characteristics. This enables fair comparisons against emerging techniques and supports multimodal intrusion detection methodologies, which combine packet payload with flow-level information. 

The CICIDS2017 dataset was created through a five-day experiment, and organized into separate daily capture files, as summarized in Table~\ref{tab:cicidsattacks}. Each file is assigned a unique ID and is characterized by its  attack scenarios and the total number of captured malicious samples.

\vspace{-12pt}
\begin{table}
\caption{Table captions should be placed above the
tables.}\label{tab:cicidsattacks}
\begin{tabular}{|l|l|p{4.5cm}|p{2.5cm}|}
\hline
ID& File & Attacks &Total Malicious Samples\\
\hline
 -& Monday-WorkingHours&Only benign &- \\
 1& Tuesday-WorkingHours&FTP-Patator, SSH-Patator &13835\\
 2& Wednesday-WorkingHours&DoS GoldenEye, DoS Hulk, DoS Slowhttptest, DoS Slowloris, Heartbleed  &252672\\
 3& Thursday Morning&Web Attacks including: Brute Force, XSS, SQL Injection &2180 \\
 4& Thursday Afternoon&Infiltration &36\\
 5& Friday Morning&Botnet  &1966 \\
 6& Friday Afternoon Port Scan&Port Scan &158930\\
 7& Friday Afternoon DDOS LOIT&DDoS &128027\\
\hline
\end{tabular}
\end{table}

\vspace{-12pt}

To examine a model's generalization capabilities on unseen attacks, the daily-file split structure of the CICIDS2017 dataset is exploited, in which each daily file serves as an independent dataset for analysis.  Because the Monday file of the CICIDS2017 dataset includes only benign data, it is excluded from our experimental study.

This setup leverages temporal variation, while also keeping the network environment invariant, allowing the generalization test to unseen attacks without domain shifts and feature constraints. Unlike the Leave-One-Attack-Out protocols adopted by some studies, the presented experimental setup is more broadly representative and generic. To the best of our knowledge, this cross-day intra-dataset generalization evaluation has not been explored in the literature, offering a high-level model's generalization evaluation without introducing domain shifts. 

\subsection{Experimental Design}
\label{subsec:design}

In an effort to examine the dimensionality parameter in the generalization behavior of classifiers, three feature selection approaches have been employed, and organized into two studies. The first study, referred to as \textit{“Experimental Study I”}, serves as a comprehensive baseline experiment, \textit{full-feature}, where no feature selection is applied, and all 77 features are retained. It isolates the classifier effect and demonstrates the generalization among distinct attacks under different sampling schemes. The second study, referred to as \textit{Experimental Study II}, introduces the remaining dimensionality reduction techniques, including a \textit{local} and a \textit{greedy global} feature selection. The second study interprets feature importance from a trained model perspective, reflecting the realistic model's viewpoint on how features are used to make predictions, unlike overly optimistic results obtained from handcrafted selection, offering a more accurate representation of feature importance.

Both testbeds share the following methodological foundation. A binary ML classifier is trained on each daily file and evaluated on the remaining day-datasets, simulating exposure to novel attack types.  Binary classification is selected to emphasize the fundamental IDS task of distinguishing malicious from benign traffic, regardless of the attack type, and therefore prioritizing attack detection over taxonomy.

Since the nature of the experiment is to highlight the importance of generalization and emphasize behavioral differences, rather than optimized generalization models, they are trained under default settings without fine-tuning. This allows for an unbiased, consistent, and fair comparison of the general behavior of classifiers. Nevertheless, it is acknowledged that this may favor architectures robust to hyperparameters. To enable reproducibility, a fixed random state of zero is applied where applicable.

To account for class imbalance in the test sets, Random Under-Sampling is applied to each test set prior to evaluation, retaining only the original data. This ensures accuracy remains a meaningful metric. Given the large number of models and configurations presented, reports include only accuracy and recall as the primary metrics. Additional metrics are omitted to avoid burdening the results of these small-scale experiments. 


\subsubsection{Experimental Study I: Full Featured Set}

On the full feature set, Study I explores the classifier capabilities and attack pattern relations. Two tree-based supervised ML classifiers are employed: Random Forest (RF), which has consistently demonstrated robust performance in recent literature~\cite{umar2025effects}, and XGBoost (XGB), which is known for its robustness in handling complex data. To address class imbalance, two sampling techniques have been utilized, namely, SMOTE to synthetically generate minority class instances and avoid original data information loss, and Random Under-Sampling to rely solely on the predictions on original observed data. Finally, two scaling techniques are tested: Min-Max scaling and Z‑score standardization. Only slight variations in performance were observed between these two normalization techniques.  A detailed version of all hyperparameters used in the experiments is presented in Table~\ref{tab:specs}. In total, each model, for each day-dataset, was evaluated across all possible configuration combinations. 

\begin{table}
\caption{Hyperparameters and specifications}
\label{tab:specs}
\centering
\begin{tabular}{|l|p{2.5cm}|p{7cm}|}
\hline
\textbf{Component} & \textbf{Option} & \textbf{Key Hyperparameters} \\
\hline
Classifier & RF & Estimators = 100, criterion = entropy, random state = 0 \\
 & XGBoost & Default hyperparameters, random state = 0 \\
\hline
Normalization & Z-score & -- \\
 & Min-Max & Feature range = (0,1) \\
\hline
Sampling & Random Under-Sampling & Sampling strategy = minority class number of samples, random state = 0 \\
 & SMOTE & Sampling strategy = auto, random state = 0 \\
\hline
Feature Selection & None & All 77 features used \\
 & Local & Top 20 based on model-specific feature importance \\
 & Global Mean & Features ranked per attack, selecting the global top 20 \\
\hline
\end{tabular}
\end{table}

\subsubsection{Experimental Study II: Feature Selection Set}

To assess the impact of feature dominance, particularly since models are not fine-tuned, and the objective is to concentrate on model-inherent behavior, feature selection is introduced in Study II. Three feature importance estimation techniques are initially considered, namely, XGBoost built-in feature importance, Permutation importance, and SHAP values. XGB feature importance measures importance based on tree splits, reflecting model usage. Permutation importance is a model-agnostic approach that evaluates a feature’s contribution by reducing predictive performance when the values are randomly permuted. Finally, SHAP \cite{lundberg2017unified}, a game-theoretic approach, assigns each feature a contribution value for individual predictions. 

SHAP is adopted in our methodology due to its widespread adoption in ML explainability literature and its ability to account for feature interactions and non-linear effects as opposed to correlation-based approaches. However, we acknowledge that SHAP reflects a model's reliance, and not feature utility across datasets. A fixed value of 20 features is selected to balance informativeness and model simplicity. A broader sweep is left for future tuning-focused work.

Two selection strategies are compared. The \textit{local} feature selection approach selects the top 20 features with the highest SHAP importance for the current training dataset. On a similar note, a \textit{global greedy} feature importance is presented to compute the global importance. The mean absolute SHAP value for each feature \(i\) is computed for each local model \(k \in \{1, \dots, N\}\) (Eq. \ref{eq:normalizedshap}). These values are then normalized to ensure comparability across models.

\vspace{-10pt}
\begin{equation}
\text{NormalizedSHAP}_i^{(k)} = \frac{\mathbb{E}[|\text{SHAP}_i^{(k)}|]}{\sum_j \mathbb{E}[|\text{SHAP}_j^{(k)}|]}
\label{eq:normalizedshap}
\end{equation}

For each feature, the normalized values of all N local models are accumulated to derive a global ranking feature vector (Eq. \ref{eq:accumulated}), from which the identical 20 highest-ranked features are selected for all test sets.

\begin{equation}
\text{AccumulatedSHAP}_i = \sum_{k=1}^{N} \text{NormalizedSHAP}_i^{(k)} 
\label{eq:accumulated}
\end{equation}

The contrast between local and global feature selection seeks to determine whether models benefit more from task-specific discriminative features or from a universally consistent subset that captures a broader range of behavior. The classifier configurations remain consistent with the previously identified optimal setup, utilizing XGB with Random Under-Sampling and Z-score standardization.

\section{Results}
\label{sec:results}

\subsection{Experimental Study I}

In order to test the generalization across a variety of network attack scenarios, an end-to-end experimental pipeline is constructed. The framework comprises cross-day experiments that incorporate various sampling methods and feature preprocessing techniques. Each model was trained on a specific day-dataset and tested on others to evaluate its performance while exposed to unseen attack classes. Unseen attacks are defined as those not present in the training distribution, thus representing realistic, previously unencountered threats in a real production environment. 

\begin{figure}[h!]
\centering
\begin{tikzpicture}
\begin{axis}[
    width=\linewidth,
    height=5.5cm,
    enlargelimits=false,
    xlabel={(Train, Test) Dataset Pair},
    ylabel={Accuracy (\%)},
    xtick=data,
    xticklabel style={rotate=90, anchor=east, font=\scriptsize},
    xticklabels={
        {(1,1)}, {(1,2)}, {(1,3)}, {(1,4)}, {(1,5)}, {(1,6)}, {(1,7)},
        {(2,1)}, {(2,2)}, {(2,3)}, {(2,4)}, {(2,5)}, {(2,6)}, {(2,7)},
        {(3,1)}, {(3,2)}, {(3,3)}, {(3,4)}, {(3,5)}, {(3,6)}, {(3,7)},
        {(4,1)}, {(4,2)}, {(4,3)}, {(4,4)}, {(4,5)}, {(4,6)}, {(4,7)},
        {(5,1)}, {(5,2)}, {(5,3)}, {(5,4)}, {(5,5)}, {(5,6)}, {(5,7)},
        {(6,1)}, {(6,2)}, {(6,3)}, {(6,4)}, {(6,5)}, {(6,6)}, {(6,7)},
        {(7,1)}, {(7,2)}, {(7,3)}, {(7,4)}, {(7,5)}, {(7,6)}, {(7,7)}
    },
    legend style={at={(0.5,-0.2)}, anchor=north, legend columns=2},
    ymin=39, ymax=100,
    ymajorgrids=true,
    grid style=dashed
]

\addplot+[mark=*, color=blue] coordinates {
    (0,100.00) (1,50.00) (2,50.00) (3,50.00) (4,50.00) (5,50.01) (6,50.00)
    (7,49.57) (8,99.95) (9,51.90) (10,50.00) (11,49.90) (12,50.02) (13,81.73)
    (14,58.53) (15,50.11) (16,99.99) (17,50.00) (18,49.97) (19,50.01) (20,49.90)
    (21,50.00) (22,50.00) (23,50.00) (24,100.00) (25,50.00) (26,50.00) (27,50.00)
    (28,50.00) (29,49.97) (30,49.95) (31,50.00) (32,99.97) (33,49.99) (34,49.87)
    (35,50.00) (36,52.52) (37,50.67) (38,50.00) (39,50.00) (40,99.99) (41,79.22)
    (42,50.00) (43,50.01) (44,50.00) (45,50.00) (46,50.00) (47,50.01) (48,99.99)
};
\addlegendentry{RF + SMOTE}

\addplot+[mark=square*, color=red] coordinates {
    (0,100.00) (1,50.00) (2,50.37) (3,50.00) (4,50.00) (5,50.07) (6,49.99)
    (7,49.57) (8,99.95) (9,52.27) (10,50.00) (11,49.92) (12,50.01) (13,81.70)
    (14,65.84) (15,51.73) (16,99.54) (17,50.00) (18,49.59) (19,49.79) (20,49.60)
    (21,49.14) (22,48.65) (23,48.72) (24,88.89) (25,49.57) (26,48.64) (27,41.20)
    (28,50.00) (29,49.35) (30,48.94) (31,48.61) (32,99.08) (33,50.73) (34,48.86)
    (35,50.10) (36,52.35) (37,50.85) (38,50.00) (39,50.00) (40,99.99) (41,71.93)
    (42,50.00) (43,50.01) (44,50.00) (45,50.00) (46,50.00) (47,50.01) (48,99.99)
};
\addlegendentry{RF + RUS}

\addplot+[mark=triangle*, color=green!60!black] coordinates {
    (0,100.00) (1,50.00) (2,50.00) (3,50.00) (4,50.00) (5,50.20) (6,50.00)
    (7,56.25) (8,99.98) (9,93.17) (10,50.00) (11,49.95) (12,50.03) (13,81.45)
    (14,50.20) (15,50.79) (16,100.00) (17,50.00) (18,49.97) (19,50.03) (20,49.99)
    (21,50.00) (22,50.00) (23,50.00) (24,100.00) (25,50.00) (26,50.01) (27,50.00)
    (28,49.82) (29,49.96) (30,49.93) (31,50.00) (32,99.95) (33,50.80) (34,49.88)
    (35,50.05) (36,56.12) (37,57.36) (38,40.28) (39,49.97) (40,99.99) (41,72.10)
    (42,50.00) (43,50.35) (44,50.00) (45,50.00) (46,50.00) (47,50.00) (48,99.99)
};
\addlegendentry{XGB + SMOTE}

\addplot+[mark=diamond*, color=orange] coordinates {
    (0,99.99) (1,49.98) (2,50.00) (3,50.00) (4,50.00) (5,50.67) (6,49.98)
    (7,49.91) (8,99.98) (9,93.60) (10,50.00) (11,49.95) (12,50.06) (13,81.48)
    (14,60.21) (15,53.46) (16,99.72) (17,50.00) (18,49.97) (19,49.97) (20,49.36)
    (21,49.73) (22,49.37) (23,49.70) (24,94.44) (25,49.80) (26,49.90) (27,44.98)
    (28,49.93) (29,49.71) (30,49.75) (31,48.61) (32,99.80) (33,69.19) (34,46.74)
    (35,49.97) (36,59.99) (37,91.93) (38,40.28) (39,50.00) (40,100.00) (41,72.12)
    (42,62.01) (43,51.59) (44,50.00) (45,50.00) (46,50.00) (47,49.99) (48,100.00)
};
\addlegendentry{XGB + RUS}

\end{axis}
\end{tikzpicture}
\caption{Accuracy across train-test pairs for different classifiers and sampling configurations.}
\label{fig:accuracy_classifier_configs}
\end{figure}
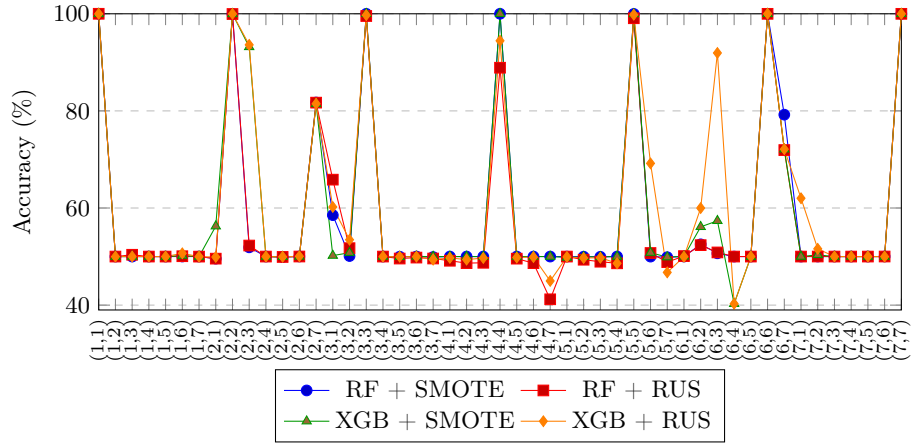

For the train and intra-day test, a 75-25 split is applied. Figure~\ref{fig:accuracy_classifier_configs} illustrates the accuracy among four configurations of RF and XGB combined with SMOTE and Random Under-Sampling.

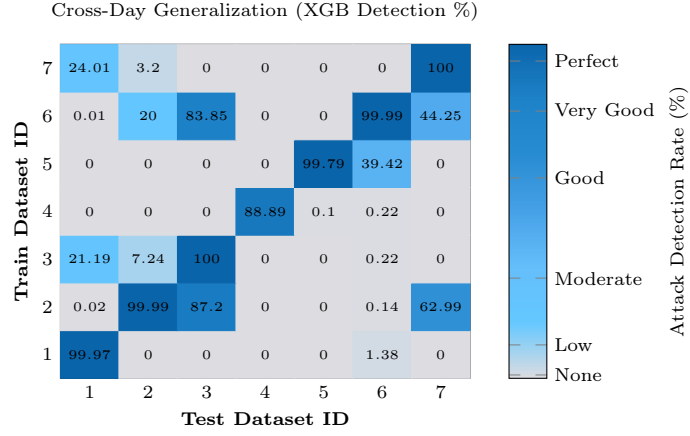
\begin{figure}[h!]
\centering
\scriptsize     
\begin{tikzpicture}
\begin{axis}[
    title={Cross-Day Generalization (XGB Detection \%)},
    xlabel={\textbf{Test Dataset ID}},
    ylabel={\textbf{Train Dataset ID}},
    xtick={1,2,3,4,5,6,7},
    ytick={1,2,3,4,5,6,7},
    enlargelimits=false,
    width=7cm,
    height=6cm,
    colormap={custom}{
        rgb255=(220,220,225) 
        rgb255=(100,200,255) 
        rgb255=(100,190,250) 
        rgb255=(70,160,230)  
        rgb255=(40,140,210)  
        rgb255=(30,130,200)  
        rgb255=(10,100,170)  
    },
    point meta min=0,
    point meta max=100,
    colorbar,
    colorbar style={
        ytick={ 1, 10, 30, 60, 80, 95},
        yticklabels={
            None,
            Low,
            Moderate,
            Good,
            Very Good,
            Perfect
        },
        ylabel={Attack Detection Rate (\%)}
    },
    nodes near coords,
    nodes near coords align={center},
    every node near coord/.append style={
        font=\tiny,
        /pgf/number format/fixed,
        /pgf/number format/precision=2
    }
]
\addplot [
    matrix plot*,
    mesh/cols=7,
    point meta=explicit
] table [meta=meta] {
x y meta
1 1 99.97
2 1 0
3 1 0
4 1 0
5 1 0
6 1 1.38
7 1 0
1 2 0.02
2 2 99.99
3 2 87.2
4 2 0
5 2 0
6 2 0.14
7 2 62.99
1 3 21.19
2 3 7.24
3 3 100
4 3 0
5 3 0
6 3 0.22
7 3 0
1 4 0
2 4 0.004
3 4 0
4 4 88.89
5 4 0.10
6 4 0.22
7 4 0
1 5 0
2 5 0
3 5 0
4 5 0
5 5 99.79
6 5 39.42
7 5 0
1 6 0.01
2 6 20
3 6 83.85
4 6 0
5 6 0
6 6 99.99
7 6 44.25
1 7 24.01
2 7 3.20
3 7 0
4 7 0
5 7 0
6 7 0
7 7 100
};
\end{axis}
\end{tikzpicture}
\caption{Cross-Day Generalization Performance of XGBoost Classifier. }
\label{fig:cross_day_xgb_heatmap}
\end{figure}

Among all tested configurations, using all features, the XGBoost classifier with Z-normalization and Random Under-Sampling demonstrated the most remarkable generalization capabilities, as depicted in Fig.~\ref{fig:cross_day_xgb_heatmap}. Nevertheless, SMOTE with XGBoost exhibited comparable prediction patterns, which are omitted for brevity. As visualized, diagonal dominance is expected due to intra-day testing. However, the off-diagonal values reveal declines in detection rates when generalization is required, especially in early-day datasets.

\subsection{Experimental Study II}

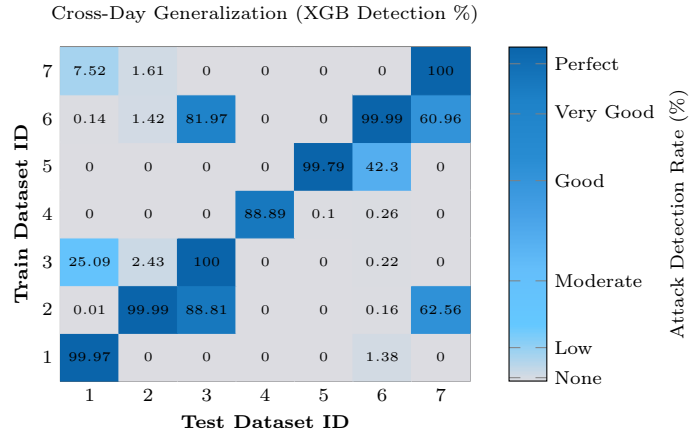
\begin{figure}[h!]
\centering
\scriptsize     
\begin{tikzpicture}
\begin{axis}[
    title={Cross-Day Generalization (XGB Detection \%)},
    xlabel={\textbf{Test Dataset ID}},
    ylabel={\textbf{Train Dataset ID}},
    xtick={1,2,3,4,5,6,7},
    ytick={1,2,3,4,5,6,7},
    enlargelimits=false,
    width=7cm,
    height=6cm,
    colormap={custom}{
        rgb255=(220,220,225) 
        rgb255=(100,200,255) 
        rgb255=(100,190,250) 
        rgb255=(70,160,230)  
        rgb255=(40,140,210)  
        rgb255=(30,130,200)  
        rgb255=(10,100,170)  
    },
    point meta min=0,
    point meta max=100,
    colorbar,
    colorbar style={
        ytick={ 1, 10, 30, 60, 80, 95},
        yticklabels={
            None,
            Low,
            Moderate,
            Good,
            Very Good,
            Perfect
        },
        ylabel={Attack Detection Rate (\%)}
    },
    nodes near coords,
    nodes near coords align={center},
    every node near coord/.append style={
        font=\tiny,
        /pgf/number format/fixed,
        /pgf/number format/precision=2
    }
]
\addplot [
    matrix plot*,
    mesh/cols=7,
    point meta=explicit
] table [meta=meta] {
x y meta
1 1 99.97
2 1 0
3 1 0
4 1 0
5 1 0
6 1 1.38
7 1 0
1 2 0.01
2 2 99.99
3 2 88.81
4 2 0
5 2 0
6 2 0.16
7 2 62.56
1 3 25.09
2 3 2.43
3 3 100
4 3 0
5 3 0
6 3 0.22
7 3 0
1 4 0
2 4 0
3 4 0
4 4 88.89
5 4 0.10
6 4 0.26
7 4 0
1 5 0
2 5 0
3 5 0
4 5 0
5 5 99.79
6 5 42.30
7 5 0
1 6 0.14
2 6 1.42
3 6 81.97
4 6 0
5 6 0
6 6 99.99
7 6 60.96
1 7 7.52
2 7 1.61
3 7 0
4 7 0
5 7 0
6 7 0
7 7 100
};
\end{axis}
\end{tikzpicture}
\caption{Cross-Day Generalization Performance of XGBoost Classifier using local feature selection for each sub-dataset}
\label{fig:cross_day_xgb_heatmap_local_feature_selection}
\end{figure}

Initially, SHAP importance is computed for each feature for all day-datasets, to perform the local feature selection for each classifier. Each day classifier is trained on the top 20 features with the highest SHAP importance, for the day-dataset employed during training. The heatmap performance, across datasets, is illustrated in Fig.~\ref{fig:cross_day_xgb_heatmap_local_feature_selection}.

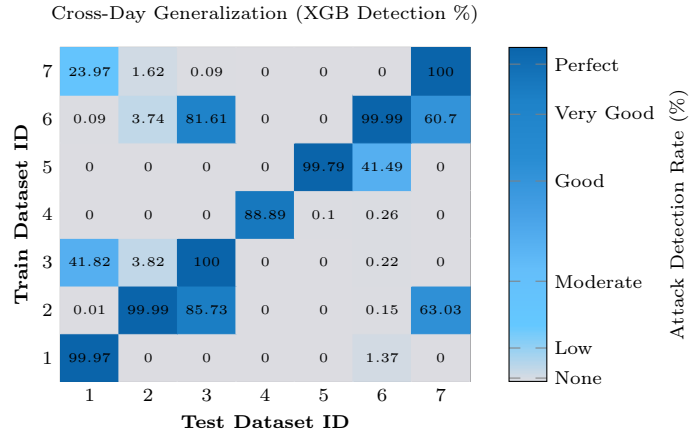
\begin{figure}[h!]
\centering
\scriptsize     
\begin{tikzpicture}
\begin{axis}[
    title={Cross-Day Generalization (XGB Detection \%)},
    xlabel={\textbf{Test Dataset ID}},
    ylabel={\textbf{Train Dataset ID}},
    xtick={1,2,3,4,5,6,7},
    ytick={1,2,3,4,5,6,7},
    enlargelimits=false,
    width=7cm,
    height=6cm,
    colormap={custom}{
        rgb255=(220,220,225) 
        rgb255=(100,200,255) 
        rgb255=(100,190,250) 
        rgb255=(70,160,230)  
        rgb255=(40,140,210)  
        rgb255=(30,130,200)  
        rgb255=(10,100,170)  
    },
    point meta min=0,
    point meta max=100,
    colorbar,
    colorbar style={
        ytick={ 1, 10, 30, 60, 80, 95},
        yticklabels={
            None,
            Low,
            Moderate,
            Good,
            Very Good,
            Perfect
        },
        ylabel={Attack Detection Rate (\%)}
    },
    nodes near coords,
    nodes near coords align={center},
    every node near coord/.append style={
        font=\tiny,
        /pgf/number format/fixed,
        /pgf/number format/precision=2
    }
]
\addplot [
    matrix plot*,
    mesh/cols=7,
    point meta=explicit
] table [meta=meta] {
x y meta
1 1 99.97
2 1 0
3 1 0
4 1 0
5 1 0
6 1 1.37
7 1 0
1 2 0.01
2 2 99.99
3 2 85.73
4 2 0
5 2 0
6 2 0.15
7 2 63.03
1 3 41.82
2 3 3.82
3 3 100
4 3 0
5 3 0
6 3 0.22
7 3 0
1 4 0
2 4 0
3 4 0
4 4 88.89
5 4 0.1
6 4 0.26
7 4 0
1 5 0
2 5 0
3 5 0
4 5 0
5 5 99.79
6 5 41.49
7 5 0
1 6 0.09
2 6 3.74
3 6 81.61
4 6 0
5 6 0
6 6 99.99
7 6 60.70
1 7 23.97
2 7 1.62
3 7 0.09
4 7 0
5 7 0
6 7 0
7 7 100
};
\end{axis}
\end{tikzpicture}
\caption{Cross-Day Generalization Performance of XGBoost Classifier using global feature selection}
\label{fig:cross_day_xgb_heatmap_global_feature_selection}
\end{figure}

This global vector highlights features that are consistently important across multiple local datasets. The results of the experiment are presented in Fig.~\ref{fig:cross_day_xgb_heatmap_global_feature_selection}. 

\begin{figure}[h!]
\centering
\scriptsize   
\begin{tikzpicture}
\begin{axis}[
    ybar,
    bar width=14pt,
    ymin=0, ymax=1.0,
    ytick={0,0.2,0.4,0.6,0.8,1.0},
    ylabel={Spearman~$\rho$},
    xlabel={Dataset},
    xtick=data,
    xticklabels={
        Tue--Brute Force,
        Wed--DoS,
        Thu--AM Web,
        Thu--PM Infl,
        Fri--AM Bot,
        Fri--PM Scan,
        Fri--PM DDoS
    },
    x tick label style={rotate=35, anchor=east},
    enlarge x limits=0.05,
    enlarge y limits=false,
    axis x line*=bottom,
    axis y line*=left,
    ymajorgrids=true,
    grid style={dashed, draw=gray!40},
    nodes near coords,
    nodes near coords align={above},
    every node near coord/.append style={
        font=\small\color{seccol},
        anchor=south,
        yshift=0pt
    },
    width=\linewidth,
    height=4cm,
]

\addplot+[draw=none, fill=blue] coordinates {
    (0, 0.6039)
    (1, 0.8575)
    (2, 0.7236)
    (3, 0.4346)
    (4, 0.5159)
    (5, 0.8350)
    (6, 0.7403)
};
\end{axis}
\end{tikzpicture}
\caption{Spearman correlation ($\rho$) between local and global feature rankings across different datasets.}
    \label{fig:Spearman}
\end{figure}
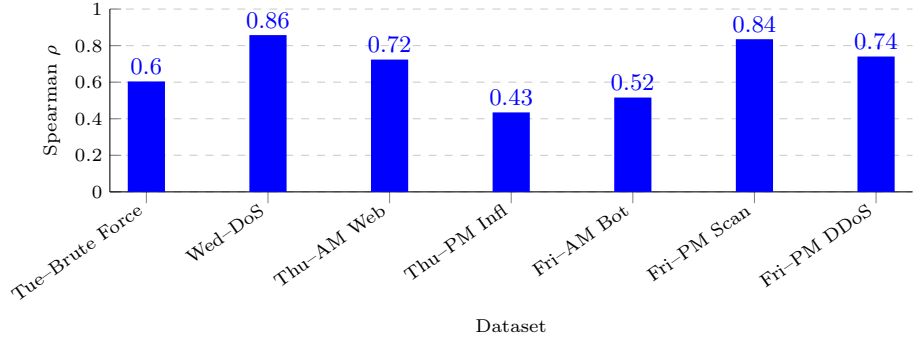

To measure the feature ranking agreement between the local and the global feature selection approaches, Spearman’s rank correlation coefficient ($\rho$) is employed. This is a non-parametric measure of the monotonic relationship between two ranked variables~\cite{hauke2011comparison}, particularly suited for measuring feature importance scores that are often non-linear. For each day-file, the Spearman correlation ($\rho$) of the local SHAP ranking and the common global SHAP importance ranking vectors is computed, as presented in Fig.~\ref{fig:Spearman}. In this setting, ($\rho$) quantifies how closely the local features reproduce the ordering of the globally aggregated importance. A high value indicates that the local most influential features align closely with the global, whereas a low value signals that the day relies on a comparatively distinctive subset. Although the global ranking is acquired from the local aggregation, it is observed that the correlations vary substantially across days.

\begin{figure}[h!]
\centering
\begin{tikzpicture}
\begin{axis}[
    width=\linewidth,
    height=5.5cm,
    enlargelimits=false,
    xlabel={(Train, Test) Dataset Pair},
    ylabel={Accuracy (\%)},
    xtick=data,
    xticklabel style={rotate=90, anchor=east, font=\scriptsize  },
    xticklabels={
        {(1,1)}, {(1,2)}, {(1,3)}, {(1,4)}, {(1,5)}, {(1,6)}, {(1,7)},
        {(2,1)}, {(2,2)}, {(2,3)}, {(2,4)}, {(2,5)}, {(2,6)}, {(2,7)},
        {(3,1)}, {(3,2)}, {(3,3)}, {(3,4)}, {(3,5)}, {(3,6)}, {(3,7)},
        {(4,1)}, {(4,2)}, {(4,3)}, {(4,4)}, {(4,5)}, {(4,6)}, {(4,7)},
        {(5,1)}, {(5,2)}, {(5,3)}, {(5,4)}, {(5,5)}, {(5,6)}, {(5,7)},
        {(6,1)}, {(6,2)}, {(6,3)}, {(6,4)}, {(6,5)}, {(6,6)}, {(6,7)},
        {(7,1)}, {(7,2)}, {(7,3)}, {(7,4)}, {(7,5)}, {(7,6)}, {(7,7)}
    },
    legend style={at={(0.5,-0.2)}, anchor=north, legend columns=3},
    ymin=39, ymax=100,
    ymajorgrids=true,
    grid style=dashed
]

\addplot+[mark=*, color=blue] coordinates {
    (0,99.99) (1,49.98) (2,50.00) (3,50.00) (4,50.00) (5,50.67) (6,49.98)
    (7,49.91) (8,99.99) (9,93.60) (10,50.00) (11,49.95) (12,50.06) (13,81.48)
    (14,60.21) (15,53.46) (16,99.93) (17,50.00) (18,49.77) (19,49.97) (20,49.36)
    (21,49.73) (22,49.37) (23,49.70) (24,98.61) (25,49.80) (26,49.90) (27,49.98) 
    (28,49.93) (29,49.71) (30,49.75) (31,48.61) (32,99.95) (33,69.19) (34,46.74)
    (35,49.97) (36,59.99) (37,91.93) (38,40.28) (39,50.00) (40,100.00) (41,72.12)
    (42,62.01) (43,51.59) (44,50.00) (45,50.00) (46,50.00) (47,49.99) (48,100.00)
};
\addlegendentry{All Features}

\addplot+[mark=square*, color=red] coordinates {
    (0,99.99) (1,49.98) (2,50.00) (3,50.00) (4,50.00) (5,50.67) (6,49.97)
    (7,49.87) (8,99.98) (9,94.40) (10,50.00) (11,49.95) (12,50.07) (13,81.27)
    (14,62.11) (15,51.07) (16,99.72) (17,50.00) (18,49.82) (19,49.96) (20,49.46)
    (21,49.76) (22,49.41) (23,49.77) (24,94.44) (25,49.85) (26,49.97) (27,44.96)
    (28,49.95) (29,49.73) (30,49.77) (31,48.61) (32,99.80) (33,70.71) (34,46.79)
    (35,50.04) (36,50.71) (37,90.99) (38,40.28) (39,50.00) (40,99.99) (41,80.48)
    (42,53.76) (43,50.79) (44,50.00) (45,50.00) (46,50.00) (47,49.99) (48,100.00)
};
\addlegendentry{Local SHAP}

\addplot+[mark=triangle*, color=green!60!black] coordinates {
   (0,99.99) (1,49.98) (2,50.00) (3,50.00) (4,50.00) (5,50.67) (6,49.98)
    (7,49.99) (8,99.98) (9,92.87) (10,50.00) (11,49.95) (12,50.06) (13,81.50)
    (14,70.47) (15,51.77) (16,99.72) (17,50.00) (18,49.82) (19,49.98) (20,49.43)
    (21,49.84) (22,49.47) (23,49.82) (24,94.44) (25,49.90) (26,50.01) (27,44.85)
    (28,49.95) (29,49.73) (30,49.72) (31,48.61) (32,99.80) (33,70.29) (34,46.79)
    (35,50.01) (36,51.87) (37,90.80) (38,40.28) (39,50.00) (40,100.00) (41,80.35)
    (42,61.98) (43,50.80) (44,50.05) (45,50.00) (46,50.00) (47,49.99) (48,100.00)
};
\addlegendentry{Global SHAP}

\end{axis}
\end{tikzpicture}
\caption{Accuracy across train-test pairs for different feature selection strategies.}
\label{fig:accuracy_feature_selection}
\end{figure}
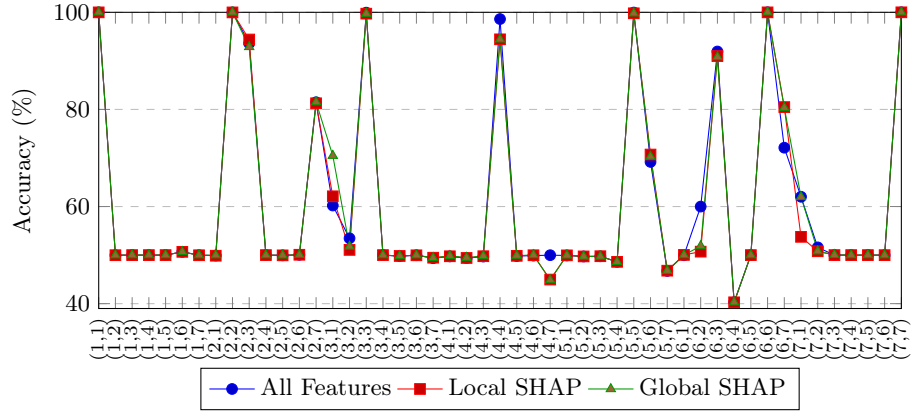

The variance in detection performance among three feature selection approaches is visualized in Fig.~\ref{fig:accuracy_feature_selection}.  It is observed that while same-day evaluations remained in optimal levels across all feature selection approaches, cross-day generalization remained challenging, especially for Local feature selection. In contrast, the Global feature selection consistently outperformed either Local SHAP or the full feature set. The results verified that global feature importance rankings represent the generalization of features across multiple attack vectors better than Local SHAP.

In conclusion, the performance comparison of the SHAP-based feature selection methods reveals that both methods, especially Local SHAP, encountered challenges with generalizability that differ in both behavior and differences in temporal activity.

\section{Results Interpretation}
\label{sec:findings}

\subsection{The Illusion of Model Efficacy}
\label{subsec:illusion}

The objective of the experimental component is not to perform optimal detection, but to explore whether the models could identify even a fraction of unseen attacks. Detecting even a small portion of novel attacks is practically valuable, demonstrating a meaningful potential to generalize models. However, while intra-day attack detection remains optimal across all classifier configurations, its effectiveness is not reflected when confronted with novel attacks, revealing significant generalization limitations. In near-zero attack detection rate test cases, beyond classifiers' inability to identify any malicious samples, it also suffers from a high amount of false negatives, resulting in the overall accuracy falling below 50\%.

The experimental setup also involved a test case scenario for each configuration, \textit{appending all daily-file datasets into a single unified dataset}, using a 75-25 train-test split. Applying the presented configuration achieved near-perfect accuracy (99.92\%), in a ten-fold cross-validation with minimal variability across folds (SD = 0.01\%). 

The absence of generalization, which is frequently overlooked in literature, is not reflected in the holistic model, highlighting an underlying risk: \textit{the illusion of model efficacy}. Assuming a consistent labeling quality and data preprocessing across days, this cross-day performance reduction is interpreted as a generalization gap. This gap suggests that models fail to retain transferable representations, which is critical for identifying new or evolving threats.

\subsection{Cross-Attack Generalization }
\label{subsec:crossattack}

In real-world scenarios, the ability of a model to detect previously unseen attacks is paramount. Despite the limitations on cross-day generalization, several specific test cases revealed promising potential. The classifier trained on DoS day-dataset achieved 93.6\% accuracy when tested on Web Attack traffic, and a reasonable 81.48\% accuracy when detecting DDoS attacks, since both DoS and DDoS attacks share structural similarities. The Web Attacks-trained classifier detected Brute Force with 60.21\% accuracy, a plausible generalization since Web Attacks include Brute Force attacks. The bot traffic trained classifier led to 69.19\% accuracy when identifying PortScan. Furthermore, the PortScan trained classifier, generalized on WebAttacks (91.93\%), DoS (60.23\%), and DDoS(72.12\%). Finally, the DDoS-trained model identified 24.01\% of Brute Force attacks, achieving a  62.01\% accuracy.

While several cross-day generalization test cases align with known structural and behavioral similarities, some test case performances may rely on normal traffic, biased features, or overlapping behavior. Nevertheless, this does not blur the generalization boundaries presented by our experimental results, and does not compromise the classifier's potential to generalize its predictions. While classifiers can generalize across distinct attack types, this behavior is inconsistent and not guaranteed.
 
\subsection{Asymmetric Generalization }
\label{subsec:asymmetric}

A more detailed study of these discrepancies leads to a notable observation: generalization between attack types tends to be asymmetric. This one-directional generalization suggests that models overfit attack-specific features rather than learning more universal behavioral patterns.  A classifier trained on Web Attacks detects a portion of Brute Force attacks, whereas the converse does not hold. A similar directional imbalance is observed between PortScan and DDoS: a classifier trained on PortScan successfully identifies DDoS, whereas DDoS-trained classifiers are ineffective in detecting PortScan. Even between structurally similar attacks, such as DoS and DDoS, one-way generalization is observed.

The asymmetric performance reflects the significance of constructing models around generalized behavioral rules, rather than relying on overall performance and best metric scores. In real-world case scenarios, where novel attacks emerge unpredictably, this asymmetry emphasizes the necessity for designing classifiers with explicit attention to generalization. A model design aimed at maximizing coverage and built on a symmetric cross-attack detection methodology may cover blind spots in practical detection, making it capable of identifying novel attacks.  

\subsection{Feature Selection Impact}
\label{subsec:features}

While feature selection can improve model efficiency, our results suggest it may inadvertently constrain generalization, highlighting the trade-off between performance and adaptability. 

 The poor \textit{Local} feature selection generalization performance reflects the underlying danger of relying on static, top-N feature selection, which is a common approach used in the literature. Such approaches tend to create models based on the specifics of known attacks, which affect the detection of new emerging threats. New feature selection mechanisms should be explored in favor of promoting more generalization and maintaining interpretability. 
 
Our findings align with concept drift and within-domain attack generalization challenges. Even under the consistent structure of CICIDS2017, the introduction of a new traffic degrades model performance. While high same-day accuracy is reported, cross-day evaluation reveals substantial performance drops, highlighting critical gaps in generalization across attacks, reflecting distributional fragility.

\section{Evaluation Criteria for ML-based Intrusion Detection Research}
\label{sec:evaluation}

The experimental findings underscore a structural weakness in ML-based IDS research and evaluation techniques. High accuracy on pooled or in-distribution datasets is often reported as evidence of a model's predictive power, thereby perpetuating the perception of an optimal model. However, such reported metrics may coexist with underlying failures in recognizing different network traffic or attack families. Such recurring methodological pitfalls in the application of Machine Learning in security research have recently been reported~\cite{arp2022and}.

In our research study, we translate the blind spots reported in the literature and the findings derived from our experimental performance collapse into a systematic set of evaluation criteria to bridge the gap between reported research performance and real-world effectiveness. Each of the presented criteria is accompanied by a reporting requirement, which transforms the observation failures into quantifiable evidence.


\paragraph{\textbf{Criterion 1: Cross-dataset Generalization Testing}}
A machine-learning-based IDS performance should be evaluated on unseen network-traffic distributions, spanning network configurations that differ from the one it was trained on. By changing the domain through cross-dataset evaluation methods, the benign baseline, the network topology, and the attack perception shift together, as when a detector trained on one network is deployed on another. Evaluation confined to in-distribution splits inflates reported metrics and conceals fragile blind spots. Research studies should therefore report a cross-dataset and cross-domain evaluation matrix including all present attacks to demonstrate the generalization and domain-shifting detection capabilities of the proposed IDS architecture.

\paragraph{\textbf{Criterion 2: Novel and Zero-day Attack Detection}}

A study should explicitly define and test against attacks that are not present in the training distribution, conducted in a fixed network environment. In this setup, the network configurations and benign traffic are maintained constant, while the adversary shifts with the introduction of an unseen attack. As demonstrated in our experimental findings, the inability to detect unseen attacks, even by models that may look optimal on the surface, achieving 99.92\% accuracy (Subsection~\ref{subsec:illusion}), and the asymmetric behavior (Subsection~\ref{subsec:asymmetric}), establish that a model's ability to detect a novel attack cannot be assumed. Evaluation of an Intrusion Detection System should adopt a Leave-One-Attack-Out protocol, excluding one attack family at a time, instead of implicitly assuming any novelty detection capability. The per-excluded-class recall should be reported, rather than
aggregating metrics that fold novel and known attack performance into a single score metric.

\paragraph{\textbf{Criterion 3: Explainability and Interpretability}}

The majority of Machine Learning-based IDS literature optimizes exclusively for detection metrics, treating the models as black boxes. Interpretability does not solely serve for transparency, but also carries operational utility. Feature-level categorizations can be translated into human-readable signatures for well-established defense tools such as Snort, Suricata, or Zeek, as well as reveal common and generalizable patterns on attack characteristics. An IDS should report which features drive its decisions, thus allowing human analysts' judgment on learned decision-making patterns. Only a handful of recent studies enable the automatic derivation of human-readable signatures, connecting ML outputs to actionable rules. Studies that report strong metrics without giving them the appropriate weight on model behavior should be regarded as incomplete, since the absence of interpretability restricts future deployments and scientific scrutiny.

\paragraph{\textbf{Criterion 4: Reproducibility and Fair Comparison}}

Results should be reproducible and obtained under clearly defined and consistent conditions. Reproducibility extends from publicly accessible sources to dimensionality reduction configurations, hyperparameters, and random seeds.  It also extends to dataset versions and the evaluation of data integrity. The evaluation on a single or outdated dataset, such as KDD99, without justification, should be considered insufficient. Many datasets can be used, ranging from relatively new benchmarks like the CICIDS2017 dataset to newly acquired with more detailed network traffic captures. This serves for both comparisons with previous work and as a standard and progressive baseline for future work. 


\paragraph{\textbf{Criterion 5: Adversarial Robustness}}
Machine Learning classifiers are vulnerable to adversarial perturbations and crafted inputs designed to evade detection. An evaluation framework should include robustness testing against evasion and poisoning techniques. Much of the existing literature assumes white-box adversaries, with complete knowledge of the detection pipeline, models, training data, and decisions, assumptions that rarely hold in real-world networks. Evaluation should reflect the constrained capabilities of a realistic adversary. Given that interpretability mechanisms may also help attackers in identifying the most critical features to perturb~\cite{chung2026deepshap}, the resilience of Machine Learning-based NIDS to adversarial strategies should be measured and reported.

\paragraph{\textbf{Criterion 6: Feature Selection Beyond Metric Improvements}}

Despite an extensive number of studies dedicated to dimensionality reduction and its decision-making importance~\cite{abdulhammed2019features}, the effect of feature selection in generalization is often overlooked~\cite{wong2025intrusion}. As reported in Subsection~\ref{subsec:features}, many studies ground their feature selection methodology on static top-N feature sets that overfit to known attacks and widen the generalization blind spots. Research studies should adopt attack-agnostic feature selection techniques, supporting deployment in heterogeneous environments, and report any in-distribution detection reduction as a tradeoff against improved generalization.

\paragraph{\textbf{Criterion 7: Multimodal Detection Techniques}}

Evaluation should consider both network flow-level features and raw packet payloads, since reliance on a single data format limits coverage. Flow-level features are susceptible to adversarial manipulation~\cite{kim2025payload}, whereas payload inspection captures semantic attack contents. Therefore, a flow-based classifier is vulnerable to evasions that a payload-aware approach might resist. Datasets such as CICIDS2017 support both flow and raw packet payloads, yet most works exploit only one. Approaches should incorporate both payloads and flow statistics, either by mapping PCAP files to existing flows~\cite{kiflay2024network}, or by further research for standardized enriched multimodal datasets~\cite{wali2024meta}.

The proposed evaluation criteria are summarized in Table~\ref{tab:summary_of_criteria}. For each criterion, three parameters are specified: (i) the information a study should disclose to fulfill it (Column 2: "Reporting Requirement"), (ii) the experimental component that tests the affiliated criterion and produces that reporting (Column 3: "Testing Protocol"), and (iii) the findings of our experimental component operating as a motive (Column 4: "Supporting Evidence"). Criteria without an entry in the last column are not directly visible from our experimental component, and are drawn from well-evidenced literature gaps.

\begin{table}
\caption{The evaluation criteria, reporting artifacts, and motivation evidence}
\label{tab:summary_of_criteria}
\centering
\begin{tabular}{@{}>{\raggedright\arraybackslash}p{0.23\linewidth} >{\raggedright\arraybackslash}p{0.27\linewidth} >{\raggedright\arraybackslash}p{0.27\linewidth} >{\raggedright\arraybackslash}p{0.20\linewidth}@{}}
\hline
\textbf{Criterion} & \textbf{Reporting \newline Requirement} & \textbf{Testing Protocol} & \textbf{Supporting Evidence}\\
\hline
1. Generalization & Cross-dataset / Environmental shift reports &  Train-on-one, test-on-other matrix & In-distribution gap (Subsec.~\ref{subsec:illusion})\\
\addlinespace
2. Zero-day detection&Intra-dataset / Hold-out attacks: per-class recall & Leave-One-Attack-Out&Asymmetric Generalization (Subsec.~\ref{subsec:asymmetric}) \\
\addlinespace
3. Explainability&Which features construct decision boundaries. Signature mapping& Feature attribution. Rule extraction& Attack patterns (Subsec.~\ref{subsec:features})\\
\addlinespace
4. Reproducibility and Fair Comparison&Dataset version, seeds, hyperparameters&Validity and Parameters matrix. Public source code&-\\
\addlinespace
5. Adversarial Robustness&Stated threat-model: Evasion resilience& Realistic adversary evaluation&-\\
\addlinespace
6. Feature generalization&Trade-off between accuracy and transferability&Attack-agnostic selection& Static top-N effect (Subsec.~\ref{subsec:features})\\
\addlinespace
7. Multi modal detection&Justification of single-modality use& Flow and payload level evaluation&-\\
\hline
\end{tabular}
\end{table}



\vspace{-20pt}
\section{Conclusions}

Taking into consideration the above criteria, the machine-learning IDS literature could be reframed from slight metric improvements in a single distribution domain toward a generalized behavioral attack profile. Unifying the above techniques, which so far have been either absent or applied as stand-alone methodologies in the literature, will result in approaches that reflect generalization, novelty handling, robustness, transparency, and research continuation. None of the criteria is onerous in an IDS architecture. However, their absence permits a detector to appear effective, concealing its inability to carry out the task it is intended for, distorting the current state of academic research. The adoption of the recommended evaluation tests and methodologies would narrow the gap between reported research performance and real-world effectiveness and would render the illusion of efficacy.


%
%
%
%
 
\end{document}